\documentclass[twocolumn, notitlepage, aps, floatfix, nofootinbib]{revtex4-1}

\usepackage{amsfonts,amsmath,amssymb}
\usepackage{hyperref}
\usepackage{graphicx}

\usepackage{fancyhdr}
\begin{document}
\preprint{APS/123-QED}
\title{Manufacturing low dissipation superconducting quantum processors}

\author{Ani Nersisyan}\thanks{Equal contribution}
\author{Stefano Poletto}\thanks{Equal contribution}
\author{Nasser Alidoust}\thanks{Equal contribution}
\author{Riccardo Manenti}\thanks{Equal contribution}
\author{Russ Renzas}
\author{Cat-Vu Bui}
\author{Kim Vu}
\author{Tyler Whyland}
\author{Yuvraj Mohan}
\author{Eyob A. Sete}
\author{Sam Stanwyck}
\author{Andrew Bestwick}
\author{Matthew Reagor}
\email{Corresponding author: matt@rigetti.com}
\affiliation{Rigetti Computing, 2919 Seventh Street, Berkeley, CA 94710}

\date{\today}

\begin{abstract}
Enabling applications for solid state quantum technology will require systematically reducing noise, particularly dissipation, in these systems. Yet, when multiple decay channels are present in a system with similar weight, resolution to distinguish relatively small changes is necessary to infer improvements to noise levels. For superconducting qubits, uncontrolled variation of nominal performance makes obtaining such resolution challenging. Here, we approach this problem by investigating specific combinations of previously reported fabrication techniques on the quality of 242 thin film superconducting resonators and qubits.
Our results quantify the influence of elementary processes on dissipation at key interfaces. We report that an end-to-end optimization of the manufacturing process that integrates multiple small improvements together can produce an average ${\overline{T}_{1}=76\pm13~\mu}$s across 24 qubits with the best qubits having ${T_1\geq110~\mu}$s. Moreover, our analysis places bounds on energy decay rates for three fabrication-related loss channels present in state-of-the-art superconducting qubits. Understanding dissipation through such systematic analysis may pave the way for lower noise solid state quantum computers.
\end{abstract}

\maketitle

Quantum computation competes with energy relaxation for solid state systems. For superconducting qubits, one strategy for mitigating dissipation is minimizing the sensitivity of qubits to sources of loss through design. Highly tailored designs of transmon qubits can isolate these circuits from dielectric loss \cite{Dial2016}, dipole radiation \cite{Paik2012, Sandberg2013}, and Purcell effects \cite{Houck2008, Reed2010, Srinivasan2011, Jeffrey2014, Bronn2015}, resulting in relaxation times as long as $T_1$=160~$\mu$s \cite{Dial2016}.
Integrated quantum processors, on the other hand, must also continue to scale in computational complexity, which is not necessarily compatible with increasing amounts of isolation. 

Eliminating energy relaxation channels from processors altogether provides a path forward. Towards that end, improvements to energy relaxation times have been achieved through better fabrication methods and materials \cite{McDermott2009, Oliver2013}. 
Fabrication limits to individual qubit performance are expensive to evaluate and can vary device-to-device, and therefore, studies have historically emphasized the importance of a single device interface or fabrication step on dissipation \cite{Wenner2011, Wang2015, Calusine2018,Barends2010, Sage2011, Megrant2012, Geerlings2012, Sandberg2012,Chang2013, Quintana2014,Bruno2015,Dunsworth2017,Leonard2018}. 
The metal-substrate interface has been shown to be a dominant contributor to dielectric loss in qubits \cite{Wenner2011, Wang2015, Calusine2018}. Removing oxide from this interface through substrate surface treatment has led to coplanar waveguide resonators (CPWRs) with single photon internal quality factors in excess of one million \cite{Megrant2012,Bruno2015} and relaxation times of ${T_1 \approx 60\,\mu}$s for planar transmon qubits \cite{OBrien2018}. Similarly, high quality CPWRs have been fabricated using low loss patterning techniques, which has been attributed to the choice of etch chemistry for improved device-air interfaces \cite{Sandberg2012} or reduced dielectric sensitivity \cite{Bruno2015, Calusine2018}. Additionally, optimized contact between junctions and other circuitry has resulted in reduced loss for planar transmon qubits (${T_1\approx40-50\,\mu}$s) at this metal-metal interface \cite{Chang2013, Dunsworth2017}. A comprehensive understanding of these insights could enable higher performance quantum processors.

Here, we present an extensive study on manufacturing techniques for superconducting quantum devices. 
By characterizing 96 individual CPWRs and 146 transmon qubits with common design layouts, we demonstrate that three separate fabrication optimizations can reduce the average dissipation rate of qubits ($\overline{\gamma}=1/\overline{T}_1$) in an end-to-end manufacturing process.
Each of the corresponding improvements to $\overline{\gamma}$ are on the order of ${|\delta\overline{\gamma}|\sim10}$~kHz. 
Due to the nature of noise in these systems, each change is comparable to the standard deviation, despite the significant quantity of devices investigated in this study. Nonetheless, we show that careful integration of elementary fabrication improvements can lead to lower dissipation superconducting devices. 

\begin{figure}
\centering
\includegraphics{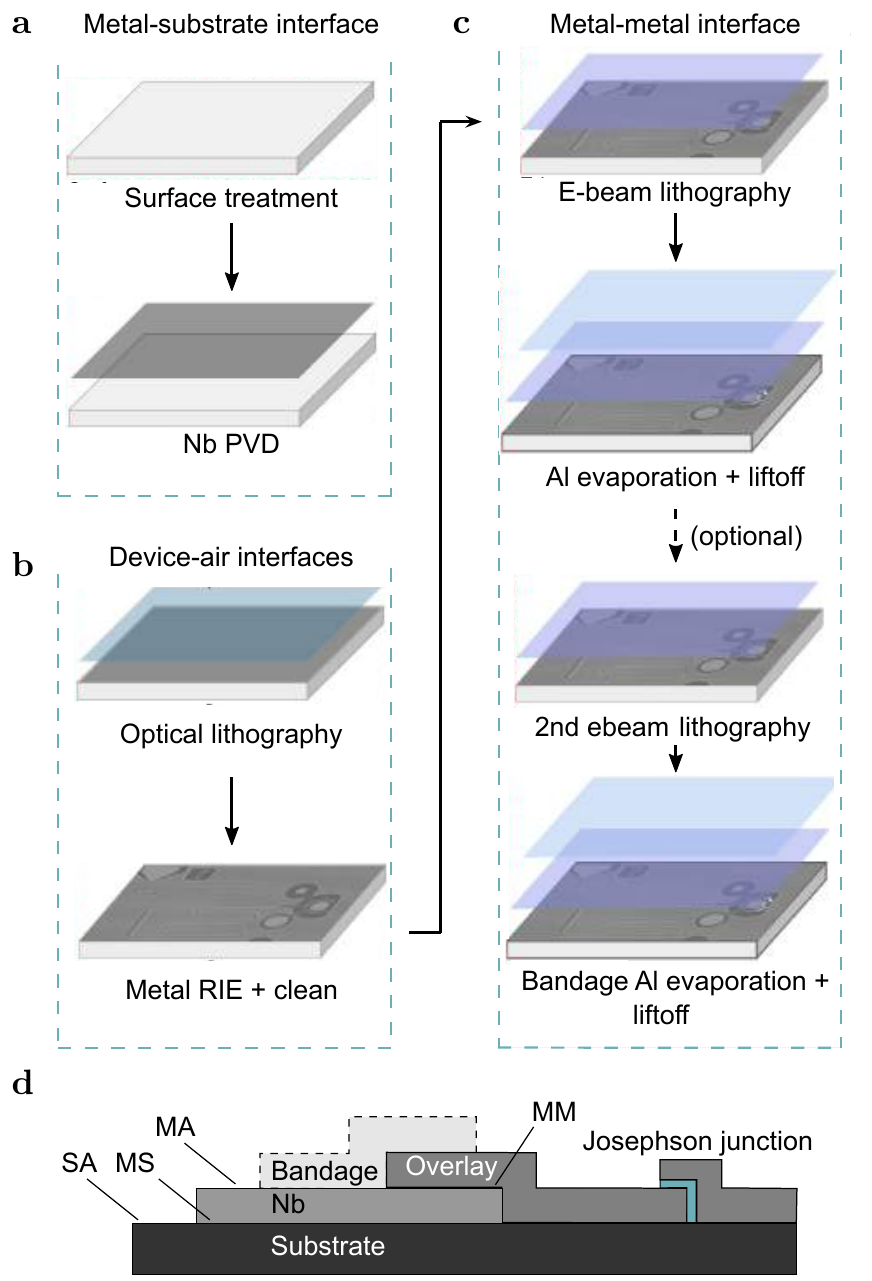}
\caption{\textbf{Fabrication flow for superconducting quantum devices.} The fabrication of superconducting qubits presented here consists of: \textbf{a,} surface treatment of the Si substrate wafer and physical vapor deposition (PVD) of the Nb film that affect the quality of the metal-substrate interface; \textbf{b,} subtractive patterning, consisting of optical lithography using positive resist (teal) and metal reactive-ion etching (RIE), followed by a cleaning step, that affect the quality of the device-air interfaces; \textbf{c,} electron beam lithography of the Josephson junctions (JJs) using MMA and PMMA resists (purple) followed by double-angle evaporation of Al (blue) and liftoff that impact the metal-metal interface. Additional e-beam lithography, Al evaporation and liftoff steps are needed for bandage qubits. \textbf{d,} Schematics of the device cross-section showing the interfaces that affect the performance of a superconducting device: metal-substrate (MS), substrate-air (SA), metal-air (MA), and metal-metal (MM).}
\label{fig:fabflow}
\end{figure}

\section{Experimental Overview}

The total energy decay rate of a resonant mode such as a qubit can be understood as the sum of decay rates through all coupled loss channels. Equivalently, if the fraction of energy stored in each lossy element is known (the participation ratios $p_i$), then the qubit decay rate can be related to an intrinsic quality (or loss tangents $\tan \delta_i$) of these elements \cite{Koch2007, Gao2008, Wenner2011, Calusine2018}, as
\begin{equation}
\gamma= \sum_i \gamma_i + \Gamma= \omega\sum_i{p_i \tan\delta_i} + \Gamma,    
\end{equation}
where $\omega/2\pi$ is the qubit mode frequency and $\Gamma$ accounts for other types of loss mechanisms, e.g. radiative decay. Each loss channel $\gamma_i$ bounds the relaxation times ($T_1 \leq 1/\gamma_i$). Improvements to $T_1$ may come from decreasing participation ratios or decreasing loss tangents, resulting in negative $\delta\gamma_i$. The goal of an end-to-end fabrication process optimization is to introduce multiple compatible improvements to loss channels, $\Delta \gamma = \sum_i \delta\gamma_i$.

In this study, we quantify interface specific losses by comparing groups of qubits that were made with identical fabrication processes, except for the steps which define a specific interface (Fig.~\ref{fig:fabflow}a-c). 
We use two device designs that are compatible with high throughput testing of internal quality factors of CPWRs (devices enumerated as Rd\# with 8 CPWR per device) or qubit relaxation times (Qd\# with 8 qubits per device). More detailed information about the designs and measurement setup are provided in Section \ref{supplement} and a list of devices is presented in Table~\ref{tab:1}. 

We first consider the influence of the metal-substrate (MS) interface (Fig.~\ref{fig:fabflow}a). We compare the control substrate surface treatment, consisting of a standard clean~\cite{Kern1990} and subsequent immersion in buffered oxide etch (BOE), against two additional treatments: Ar$^{+}$ ion milling~\cite{Geerlings2012,Megrant2012,Dunsworth2017, Vahidpour2017} and exposing to hexamethyldisilazane (HMDS) \cite{Bruno2015}. After substrate treatment, wafers are coated with Nb through physical vapor deposition (PVD), which completes the MS interface. Next, we focus on the metal-air (MA) and substrate-air (SA) interfaces defined in the subtractive patterning step (see Fig.~\ref{fig:fabflow}b).  Large device features (${\geq 10~\mu}$m) are defined with optical lithography followed by reactive-ion etching (RIE) with SF$_6$ to remove the exposed metal and achieve over-etch into silicon. After stripping the resist, the wafers are exposed to oxygen plasma ashing and again immersed in BOE solution. Finally, we investigate the metal-metal (MM) interface defined in connecting the Al Josephson junctions (JJs) to the rest of the Nb circuitry (Fig.~\ref{fig:fabflow}c). The JJ definition process consists of patterning a bilayer resist with electron beam lithography, followed by double-angle evaporation of two Al layers with a controlled oxidation in between. We study two different methods of achieving good contact transparency between Al and Nb: designing large metal overlay \cite{Chang2013,Leonard2018} or depositing an additional Al layer (the ``bandage'' layer) across the two metals \cite{Dunsworth2017}. A schematic of these layers is shown in Fig.~\ref{fig:fabflow}d and more details about the fabrication procedures are provided in Section~\ref{supplement}.

\begin{figure}[t] %htp
\centering
\includegraphics{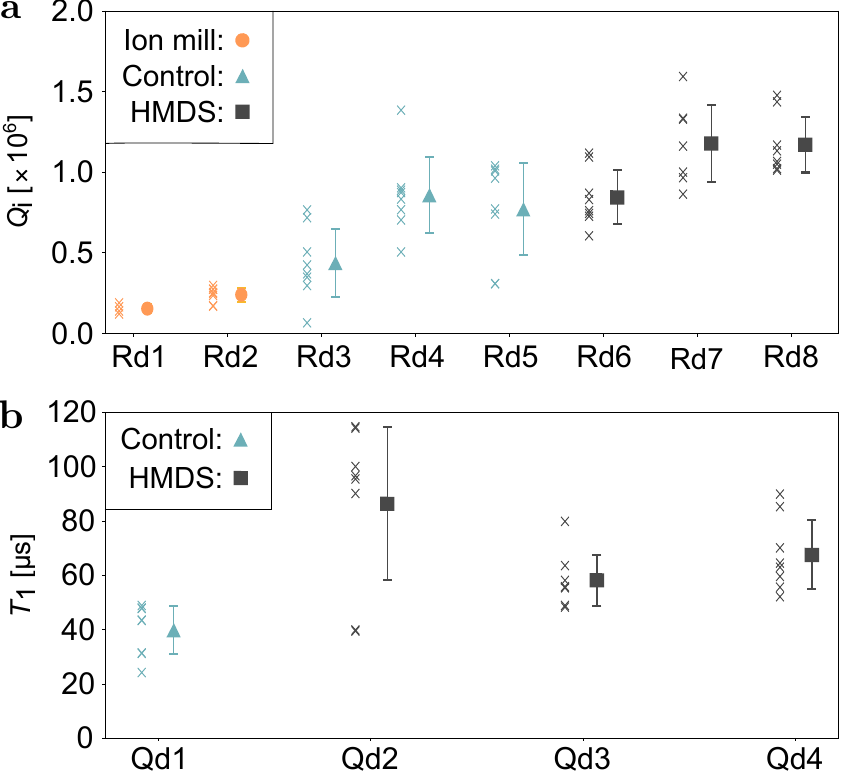}
\caption{\textbf{Performance enhancement from metal-substrate interface treatment.} \textbf{a,} Internal quality factors ($Q_\textrm{i}$) at single photon powers of resonator devices prepared using three different surface preparation methods of the silicon substrate: ion milling (orange dots), control (teal triangles), and passivation with hexamethyldisilazane (grey squares). Crosses correspond to the measured $Q_\textrm{i}$, whereas the error bars indicate standard deviations for each device. \textbf{b,} Relaxation times ($T_1$) of qubit devices prepared using similar surface treatments: control (teal triangles) and HMDS passivation (grey squares). Crosses indicate the average of $T_1$ data collected for each qubit on a device, whereas the error bars indicate thestandard deviations for each device.}
\label{fig:surface_treatment}
\end{figure}

\begin{figure*}[b] %ht
\centering
\includegraphics[scale=0.94]{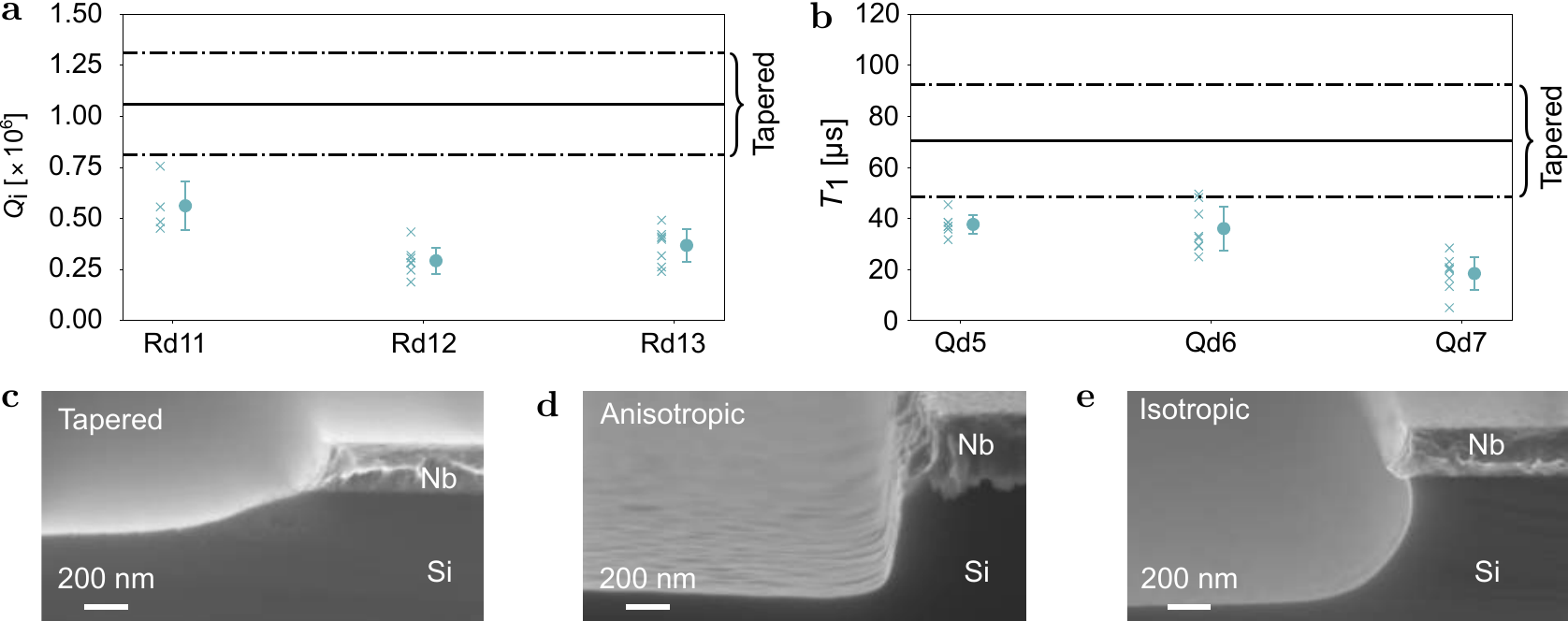}
\caption{\textbf{Performance effects from subtractive patterning etch microsctructure.} \textbf{a,} Quality factor measurements of CPWR devices with the anisotropic (\textbf{d}) profile. Results related to the tapered profile are represented by the mean of RD6, RD7 and RD8 (see Fig.~\ref{fig:surface_treatment}a) and the standard deviation of these measurements (solid and dashed lines, respectively). The anisotropic profile underperforms this mean. \textbf{b,} Results for $T_1$ of the corresponding qubit devices are consistent with the CPWR results. The tapered profile is represented by the mean of Qd2, Qd3, and Qd4 (see Fig.~\ref{fig:surface_treatment}b) and outperforms the anisotropic profile by a similar factor. \textbf{c}-\textbf{e,} Scanning electron microscope images of the cross section of the three patterning etch profiles used for this study. The dark regions at the bottom correspond to the silicon substrate, whereas the brighter regions on top show the Nb metal. The tapered profile (\textbf{c}) is our baseline profile, generated by a stop-on-silicon (timed) SF$_6$ etch process. The other two profiles are generated by an anisotropic (\textbf{d}) and isotropic (\textbf{e}) over-etch into silicon.}
\label{fig:etch_profile}
\end{figure*}

\begin{figure*}[ht]
\centering
\includegraphics[width=1.75\columnwidth]{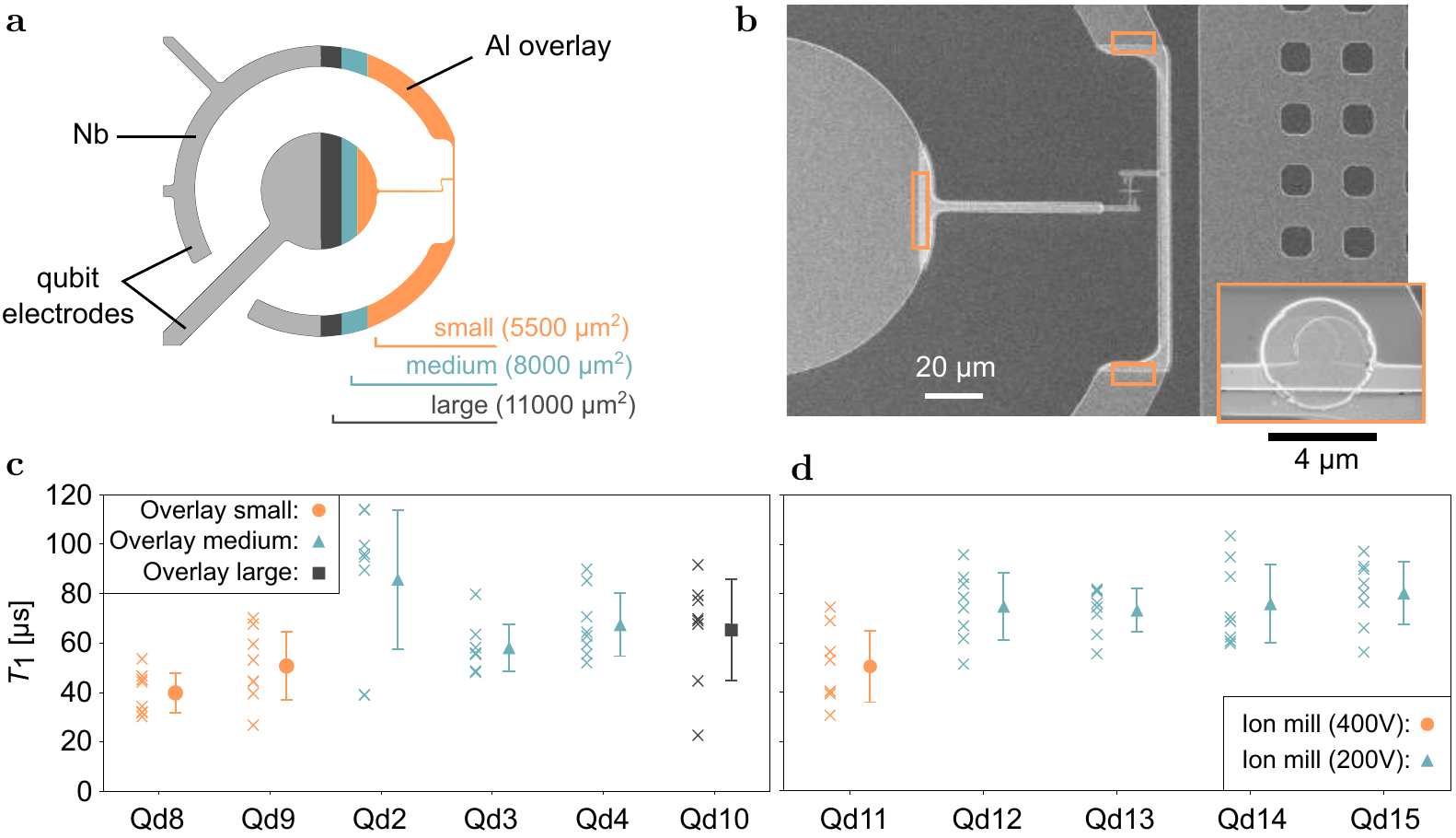}
\caption{\textbf{Relaxation time enhancement from optimized Josephson junction contacts.} \textbf{a,} Schematics of the qubit and the  three different Al overlay configurations: small (orange), medium (teal), and large (dark grey). \textbf{b,} Scanning electron microscope image of a device fabricated with the additional Al bandage layer, as highlighted in the inset. Bandage layers are applied at the locations indicated by the orange boxes. \textbf{c,} Relaxation times of qubit devices prepared without any bandage layer between the Nb device and the Al JJ leads. \textbf{d,} Relaxation times of qubit devices prepared with an additional bandage layer. Two distinct voltages are used to ion mill the metal oxide prior to the deposition of the bandage layer: $200\,$V and $400\,$V.} 
\label{fig:bandaid}
\end{figure*}

\section{Results}
\subsection{Optimizing metal-substrate interface}
\label{surface}

We begin by studying the metal-substrate (MS) interface  (Fig.~\ref{fig:fabflow}a). Preparation of the silicon surface is conducted using one of three methods: no further treatment (control), ion milling, or HMDS passivation. These treatments determine the interface below the Nb device structures, such as the CPWR and qubit capacitors. Measurements of single photon internal quality factors for resonators fabricated with each of these three methods are shown in Fig.~\ref{fig:surface_treatment}a.
Resonators from the control group (Rd3-Rd5) show ${\overline{Q}_\textrm{i}=(0.69\pm0.31)\times10^6}$, where ${\overline{Q}_\textrm{i}}$ is the arithmetic mean of the individual resonator $Q_{\textrm{i}}$ with the standard deviation over resonators. Relative to this baseline, we find that ion mill cleaning before metal deposition substantially decreases performance (${\overline{Q}_\textrm{i}=(0.15\pm0.03)\times10^6}$ from Rd1, Rd2), consistent with previous results~\cite{Wisbey2010,Geerlings2012,Dunsworth2017}. This is attributed to Si surface roughness introduced by the ion mill.
On the other hand, we find that HMDS passivation increases the internal quality factor of Nb resonators to ${\overline{Q}_\textrm{i}=(1.06\pm0.25)\times10^6}$ (devices Rd6-Rd8), consistent with observations from NbTiN on silicon resonators \cite{Bruno2015}.
These high quality factors are attributed to the hydrosilylation reaction that inhibits the formation of a native silicon-oxide layer at what becomes the MS interface \cite{Bruno2015,Fenner1989}, reducing the dielectric loss at this interface.

To illustrate the effect on qubits, we compare the relaxation times of a representative subset of qubits made with the control process versus HMDS passivation (Fig.~\ref{fig:surface_treatment}b). We find ${\overline{T}^\textrm{C}_1=40\pm9\,\mu}$s for eight qubits from the control group (Qd1) and  ${\overline{T}^\textrm{H}_1=70\,\pm\,22\,\mu}$s for twenty four qubits from the HMDS group (Qd2-Qd4), where $\overline{T}_1$ is the arithmetic mean of the individual qubit $T_{1}$ with the standard deviation over qubits. Therefore, similar to the effect on resonator quality factors, HMDS passivation improves the relaxation times of qubits.

We infer a lower limit to the change in the thickness of the oxide of the MS interface (assuming a constant value for $\tan\delta_{\textrm{MS}}$), by comparing the decay rate of three groups of control qubits (Qd1, Qd16-Qd18, and Qd19-Qd20) to three groups of HMDS-treated qubits (Qd2-Qd4, Qd11, and Qd12-Qd15). Since all other fabrication steps were kept the same, we calculate the change in the MS decay rate as the weighted average of three separate $\delta\overline{\gamma}_{\textrm{MS}}^{\textit{i}}$ obtaining $\delta\overline{\gamma}_{\textrm{MS}}=-(9\pm4)$~kHz. We attribute this change to a reduction in the participation ratio $p_{\textrm{MS}}$ from reduced thickness of the passivated oxide layer between the metal and the substrate.
We estimate this change as $\delta\overline{\gamma}_\textrm{MS}^{\textit{i}}
=(\overline{\gamma}^{\textrm{H}, i}_\textrm{MS}-\overline{\gamma}^{\textrm{C}, i}_\textrm{MS})
=K\omega( t^{\textrm{H}, i}_\textrm{MS} - t^{\textrm{C}, i}_\textrm{MS})$
where $t^{\textrm{C}, i}_\textrm{MS}$ and $t^{\textrm{H}, i}_\textrm{MS}$ are the thicknesses of the MS interface for control and HMDS-treated qubits in each group, and $K$ is a factor absorbing all constants.
This is translated into a relative value by renormalizing over the control decay rate
$|\delta\overline{\gamma}_\textrm{MS}^{i}| / \overline{\gamma}^{\textrm{C}, i} \leq |\delta\overline{\gamma}_\textrm{MS}^{i}| / \overline{\gamma}^{\textrm{C}, i}_\textrm{MS}$. In order to estimate a total thickness change of the MS interface, we calculate the weighted average of each ${\delta\overline{\gamma}_{\textrm{MS}}^{\textit{i}}}$, finding
\begin{equation}
\frac{t^{\textrm{C}}_\textrm{MS}-t^{\textrm{H}}_\textrm{MS}}{t^{\textrm{C}}_\textrm{MS}}
\geq 0.37\pm 0.13.
\end{equation}
Similar thickness variations have been observed for silicon oxidation under different conditions \cite{morita1990}.

The participation ratio of the MS interface can also be reduced with isotropic etching of the silicon substrate (see Fig.~\ref{fig:etch_profile}e).
By removing approximately 400~nm of silicon, our finite element modeling suggests a 40\% reduction in the participation ratio of the metal-substrate interface, which is comparable to our estimate for the reduction of the oxide thickness due to HMDS passivation. We find that even with the control surface preparation (Rd9, Rd10 in Table~\ref{tab:1} and Fig.~\ref{fig:ba_no_hmds}a), we recover similar performance to the HMDS-treated resonator samples ($Q_\textrm{i}$=(0.74$\pm$0.19$)\times 10^6$). Even though both HMDS and isotropic etching produce a similar effect on metal-substrate participation ratio, we see that the qualities of isotropically etched samples are worse than those treated with HMDS.
We attribute this to the increased participation ratio of the metal-air interface, which has the largest tangent loss~\cite{Woods2018,Calusine2018}.

HMDS passivation and isotropic etch are two solutions to reduce the influence of the metal-substrate interface on the device performance.
However, cross-sectional studies of the isotropic etch profile indicate that integrating junctions for qubit fabrication requires non-trivial engineering, due to the necessity for the JJ metal to climb the concave sidewall. For this reason, fabricating qubits with isotropic profiles was left out of scope for this work.

\subsection{Optimizing device-air interfaces}
\label{profiles}

We next consider the effects of fabrication on dissipation at the device-air interfaces MA and SA as indicated in Fig.~\ref{fig:fabflow}d. Consistent with \cite{Sandberg2012}, we observe that the microscopic profile of the patterned structure is important for achieving long qubit relaxation times. Three types of microstructures obtained from separate etch techniques are presented here (Fig.~\ref{fig:etch_profile}c-e), all based on reactive-ion etching (SF$_6$) to remove Nb.

A highly anisotropic etch in Fig.\ref{fig:etch_profile}d is generated with the goal of reducing participation ratios \cite{Bruno2015, Gambetta2017, Calusine2018, Sandberg2012, Vissers2012} without creating a concave structure (Fig.~\ref{fig:etch_profile}d). 
Devices produced in this manner (Rd11-13, Qd5-Qd7) had quality factors of $Q_\textrm{i}=(0.38\pm0.13)\times 10^6$  and qubit relaxation times of $T_1=30\pm11\,\mu$s. For the tapered profile shown in Fig.~\ref{fig:etch_profile}c, the etching is timed to stop on Si. 
% with a thick resist protecting the Nb metal.  
The maximum over-etch into silicon measured by cross-sectional SEM is found to be less than $200\,$nm. 
This process was used to fabricate devices Rd6-Rd8, and resulted in an average CPWR quality of $\overline{Q}_\textrm{i}=(1.06\pm0.25)\times10^6$ (Fig.~\ref{fig:etch_profile}a). The tapered etch profile is well-suited for integrating JJs, where the Al may be required to climb from the over-etched Si to the top of the Nb. 

We compare the relaxation times of qubits with tapered and anisotropic etch profiles (Qd2-Qd4 and Qd5-Qd7, respectively) which are otherwise identically fabricated. As shown in Fig.~\ref{fig:etch_profile}b, the qubits with tapered profile outperform those with anisotropic profile having $\overline{T}_1=70\pm22\,\mu$s as indicated by the horizontal lines. Following the analysis in the previous section, we estimate a reduction of $\delta \overline{\gamma}_\textrm{XA}= -(19\pm13)$~kHz, where XA indicates combined MA and SA. Note that while the standard error in this estimate is large, we also find that 92\% of the qubits with the tapered profile have $T_1$ greater than or equal to that of the best qubits from the anisotropic group. We attribute this difference to the roughened Nb edges produced by the more aggressive RIE for the anisotropic etch, which are more likely to host two-level system (TLS) defects \cite{Martinis2005}.
Furthermore, we observe a weaker internal quality factor power dependence for trenched than for tappered resonators, suggesting a different leading-order loss mechanism for the two.

\subsection{Optimizing metal-metal interface}
\label{bandaid}

In our final optimization, we consider loss at the metal-metal interface between Josephson junctions and the rest of the circuitry (Fig.~\ref{fig:fabflow}c). Importantly, to make the previously discussed optimizations compatible with functional qubits, electrical contact must be established between subtractively patterned Nb metal and Josephson junctions that are defined in a separate Al liftoff step. Finite contact resistance at the resulting metal-metal interface can result in Ohmic dissipation for qubits. To investigate this loss channel, we establish connection between Al/Nb metals using two primary approaches. In the first, we vary the dimension of an Al metal layer that overlays Nb (Fig.~\ref{fig:bandaid}a). This Al metal is defined simultaneously with the double-angle junction deposition, which includes \textit{in situ} ion milling ($160\,$V and $10\,$mA) prior to deposition. Furthermore, the Al overlay metal also forms the leads of the junction. The second approach to connecting Al/Nb is similar to the ``bandage'' procedure reported in Ref. \cite{Dunsworth2017} (Fig.~\ref{fig:bandaid}b). In this method, a second electron beam lithography step exposes a small area at the Al-Nb border where an \textit{in situ} ion milling step at high power ($200-400\,$V and $30\,$mA) removes the Nb oxide before the deposition of an Al bandage layer. The localized ion milling minimizes the damage to the Nb and Si surfaces. The locations of the bandage layers for these qubits are indicated by the orange rectangles in Fig.~\ref{fig:bandaid}b.

\begin{table}
\caption{\label{tab:1} Device fabrication parameters grouped by interface type. As explained in the main text, $\overline{T}_1$ ($\overline{Q}_i$) is the mean value of the relaxation time (internal quality factor) averaged across a device with eight qubits (resonators). Qubit (resonator) frequency is in the range 3.8 - 4.2 GHz (5.2 - 5.6 GHz).
Legend: M (metal), S (substrate), X (either M or S), A (air), I (ion mill), C (control), H (HMDS), Ta (tapered), Is (isotropic), An (anisotropic), SO, MO, LO (small, medium, large overlay).~}

\begin{ruledtabular}
\begin{tabular}{cccccc}
Device & M-S & X-A & M-M & $\overline{Q}_\textrm{i}$ [$\times10^6$] & $\overline{T}_1$ [$\mu$s] \\
\hline
Rd1\footnotemark[1] & I & Ta & - & 0.15 $\pm$ 0.03 & - \\ 
Rd2 & I & Ta & - & 0.24 $\pm$ 0.05 & -  \\
\hline
Rd9 & C & Is & - & 0.66 $\pm$ 0.15 & - \\
Rd10 & C & Is & - & 0.81 $\pm$ 0.21 & -  \\
\hline
Rd3 & C & Ta & - & 0.44 $\pm$ 0.23 & - \\
Rd4 & C & Ta & - & 0.86 $\pm$ 0.25 & -  \\
Rd5 & C & Ta & - & 0.77 $\pm$ 0.31 & -  \\
Qd1 & C & Ta & MO & - & 40 $\pm$ 10  \\
Qd16 & C & Ta & 400 V & - & 33	$\pm$ 10 \\
Qd17 & C & Ta & 400 V & - & 37 $\pm$ 13 \\
Qd18\footnotemark[1] & C & Ta & 400 V & - & 39	$\pm$ 4 \\
Qd19 & C & Ta & 200 V & - & 45	$\pm$ 15 \\
Qd20 & C & Ta & 200 V & - & 52	$\pm$ 10 \\
\hline
Rd6 & H & Ta & - & 0.84 $\pm$ 0.18 & - \\
Rd7 & H & Ta & - & 1.18 $\pm$ 0.26 & -  \\
Rd8 & H & Ta & - & 1.17 $\pm$ 0.18 & -  \\
Qd2 & H & Ta & MO & - & 86 $\pm$ 30 \\
Qd3 & H & Ta & MO & - & 58 $\pm$ 10  \\
Qd4 & H & Ta & MO & - & 68 $\pm$ 14  \\
Qd8 & H & Ta & SO & - & 39 $\pm$ 9 \\
Qd9 & H & Ta & SO & - & 48 $\pm$ 14  \\
Qd10 & H & Ta & LO & - & 64 $\pm$ 23  \\
Qd11 & H & Ta & 400 V & - & 52 $\pm$ 16  \\
Qd12 & H & Ta & 200 V & - & 75 $\pm$ 16 \\
Qd13 & H & Ta & 200 V & - & 71 $\pm$ 9  \\
Qd14 & H & Ta & 200 V & - & 76 $\pm$ 17  \\
Qd15 & H & Ta & 200 V & - & 80 $\pm$ 14  \\
\hline
Rd11\footnotemark[1] & H & An & - & 0.56 $\pm$ 0.14 & - \\
Rd12 & H & An & - & 0.29 $\pm$ 0.07 & -  \\
Rd13 & H & An & - & 0.37 $\pm$ 0.08 & -  \\
Qd5\footnotemark[2] & H & An & MO & - & 38 $\pm$ 5 \\
Qd6 & H & An & MO & - & 36 $\pm$ 9  \\
Qd7 & H & An & MO & - & 19 $\pm$ 7  \\
\end{tabular}
\end{ruledtabular}
\footnotetext[1]{Four elements characterized due to experimental limitations.}
\footnotetext[2]{Six elements characterized due to finite JJ yield.}
\end{table}

The average relaxation times for devices fabricated with the overlay technique and with the bandage layer are presented in Fig.~\ref{fig:bandaid}c-d, respectively. The samples with the overlay are divided into three categories based on the size of the Al/Nb overlay: small ($5500~\mu\textrm{m}^2$), medium ($8000~\mu\textrm{m}^2$), and large ($11000~\mu\textrm{m}^2$) (geometries shown in Fig.~\ref{fig:bandaid}a). We find that the relaxation times of devices are affected by the size of the overlay. Devices with small overlay underperform the other two geometries showing $\overline{T}_1=45\pm13\mu$s, while qubits with medium and large overlays have average $\overline{T}_1=70\pm22\,\mu$s and $\overline{T}=65\pm20\,\mu$s, respectively. We note that one of the eight qubit devices with medium overlay remarkably shows an average of $\overline{T}_1=86\pm33\,\mu$s, and two of the qubits on this device each have $T_1=114\pm19\mu$s.

The samples with the bandage layer are grouped according to the ion milling voltage ($200~\textrm{V}$ or $400~\textrm{V}$). For these devices, the data indicates that ion milling Nb can negatively affect qubit relaxation times (Fig.~\ref{fig:bandaid}d), similar to results obtained for resonator devices. Devices ion milled at $200\,$V have $\overline{T}_1=76\pm13\,\mu$s, while those ion milled at $400\,$V have $\overline{T}_1=50\pm15\,\mu$s. We have obtained similar results for qubits that were not treated with HMDS (see Table~\ref{tab:1} and Fig.~\ref{fig:ba_no_hmds}b). These observations suggest that minimizing ion milling on metal can be advantageous for achieving longer qubit relaxation times.

To bound potential losses at the metal-metal interface that have been mitigated, we follow the analysis of the previous sections.  We compare the qubits with medium Al/Nb overlay to those with small overlay. The optimized Al/Nb overlay results in a change in rate of ${\delta\overline{\gamma}_{\textrm{MM}}= -(8\pm8)}$~kHz. Moreover, 83\% of the qubits from the medium overlay group are found to have longer relaxation times than the median value of the small overlay samples.

Both overlay and bandage techniques are viable for integrating JJs. By optimizing parameters such as Al/Nb overlay area and ion milling voltage, we have successfully demonstrated average qubit relaxation times as long as $70\,\mu$s using both of these methods. However, we note that the overlay method eliminates additional e-beam lithography, Al deposition, and liftoff steps. 

\section{Conclusion} 
We have shown that carefully integrating multiple improvements together represents a promising path forward for quantum devices. Our work considered the quality of over 200 individual CPWRs and qubits. These circuits were made with 14 unique combinations of fabrication techniques. By finding the manufacturing parameters that minimize loss at key interfaces, we estimate an improvement to average qubit decay rate ${\Delta\overline{\gamma}=(\delta\overline{\gamma}_\textrm{MS}+\delta\overline{\gamma}_\textrm{XA}+\delta\overline{\gamma}_\textrm{MM}}$), assuming the dissipation channel eliminated within each module is unique. For the devices in this study, our analysis finds $\Delta\overline{\gamma}=-(30\pm16)$~kHz. We infer from this value that similar superconducting qubits fabricated with the steps shown in Fig.~\ref{fig:fabflow} can be brought from average relaxation times of $\overline{T}_1\leq 1/ \Delta\overline{\gamma} \sim 30~\mu$s to achieve reproducible relaxation times of ${\overline{T}_1 >70~\mu}$s by addressing losses at interfaces. Applying this framework to modules beyond this study could result in lower noise solid state quantum computers in the future.

\section{Acknowledgements}
We acknowledge Jayss Marshall, Kamal Yadav, Prarthana Sanghani, Theo Paran, Ganesh Ramachandran, Jane Oglesby, and Keith Jackson for significant process development support at the Fab-1 facility at Rigetti Computing, as well as the members of the Rigetti Software, Hardware, Quantum, Technical Operations, and Fabrication Operations teams who enabled this work. We thank Kevin O'Brien, Blake Johnson and Marcus P. da Silva for helpful conversations. 
\textbf{Funding}: Work at the Molecular Foundry was supported by the Office of Science, Office of Basic Energy Sciences, of the U.S. Department of Energy under Contract No. DE-AC02-05CH11231. This work was funded by Rigetti \& Co Inc., dba Rigetti Computing. \textbf{Author contributions}: AN, SP, NA, RM, RR, MR designed the overall experiment and lead data analysis; CB, CV developed etch techniques; TW, YM, SM, AB developed JJ techniques; AN designed the test chips with input from EAS, RM, SP, NA, MR; EAS estimated design performance limits; TW conducted SEM analysis; AN, SP, NA, RM, MR wrote the manuscript with input from all authors. \textbf{Conflicts of interest}: All authors are, have been, or may in the future be participants in incentive stock plans at Rigetti \& Co Inc. The authors declare that they have no competing interests.
\textbf{Data and materials availability}: All data needed to
evaluate the conclusions in the paper are present in the
paper and/or the Supplementary Materials. Additional
data related to this paper may be requested from the
authors.

%\bibliographystyle{naturemag.bst}
%\bibliography{references}

\begin{thebibliography}{10}
\expandafter\ifx\csname url\endcsname\relax
  \def\url#1{\texttt{#1}}\fi
\expandafter\ifx\csname urlprefix\endcsname\relax\def\urlprefix{URL }\fi
\providecommand{\bibinfo}[2]{#2}
\providecommand{\eprint}[2][]{\url{#2}}

\bibitem{Dial2016}
\bibinfo{author}{Dial, O.} \emph{et~al.}
\newblock \bibinfo{title}{Bulk and surface loss in superconducting transmon
  qubits}.
\newblock \emph{\bibinfo{journal}{Superconductor Science and Technology}}
  \textbf{\bibinfo{volume}{29}}, \bibinfo{pages}{044001}
  (\bibinfo{year}{2016}).

\bibitem{Paik2012}
\bibinfo{author}{Paik, H.} \emph{et~al.}
\newblock \bibinfo{title}{Observation of high coherence in josephson junction
  qubits measured in a three-dimensional circuit qed architecture}.
\newblock \emph{\bibinfo{journal}{Phys. Rev. Lett.}}
  \textbf{\bibinfo{volume}{107}}, \bibinfo{pages}{240501}
  (\bibinfo{year}{2011}).

\bibitem{Sandberg2013}
\bibinfo{author}{Sandberg, M.} \emph{et~al.}
\newblock \bibinfo{title}{Radiation-suppressed superconducting quantum bit in a
  planar geometry}.
\newblock \emph{\bibinfo{journal}{Applied Physics Letters}}
  \textbf{\bibinfo{volume}{102}}, \bibinfo{pages}{072601}
  (\bibinfo{year}{2013}).

\bibitem{Houck2008}
\bibinfo{author}{Houck, A.~A.} \emph{et~al.}
\newblock \bibinfo{title}{Controlling the spontaneous emission of a
  superconducting transmon qubit}.
\newblock \emph{\bibinfo{journal}{Phys. Rev. Lett.}}
  \textbf{\bibinfo{volume}{101}}, \bibinfo{pages}{080502}
  (\bibinfo{year}{2008}).

\bibitem{Reed2010}
\bibinfo{author}{Reed, M.~D.} \emph{et~al.}
\newblock \bibinfo{title}{Fast reset and suppressing spontaneous emission of a
  superconducting qubit}.
\newblock \emph{\bibinfo{journal}{Applied Physics Letters}}
  \textbf{\bibinfo{volume}{96}}, \bibinfo{pages}{203110}
  (\bibinfo{year}{2010}).
\newblock arXiv:\eprint{https://doi.org/10.1063/1.3435463}.

\bibitem{Srinivasan2011}
\bibinfo{author}{Srinivasan, S.~J.}, \bibinfo{author}{Hoffman, A.~J.},
  \bibinfo{author}{Gambetta, J.~M.} \& \bibinfo{author}{Houck, A.~A.}
\newblock \bibinfo{title}{Tunable coupling in circuit quantum electrodynamics
  using a superconducting charge qubit with a $v$-shaped energy level diagram}.
\newblock \emph{\bibinfo{journal}{Phys. Rev. Lett.}}
  \textbf{\bibinfo{volume}{106}}, \bibinfo{pages}{083601}
  (\bibinfo{year}{2011}).

\bibitem{Jeffrey2014}
\bibinfo{author}{Jeffrey, E.} \emph{et~al.}
\newblock \bibinfo{title}{Fast accurate state measurement with superconducting
  qubits}.
\newblock \emph{\bibinfo{journal}{Phys. Rev. Lett.}}
  \textbf{\bibinfo{volume}{112}}, \bibinfo{pages}{190504}
  (\bibinfo{year}{2014}).

\bibitem{Bronn2015}
\bibinfo{author}{Bronn, N.~T.} \emph{et~al.}
\newblock \bibinfo{title}{Broadband filters for abatement of spontaneous
  emission in circuit quantum electrodynamics}.
\newblock \emph{\bibinfo{journal}{Applied Physics Letters}}
  \textbf{\bibinfo{volume}{107}}, \bibinfo{pages}{172601}
  (\bibinfo{year}{2015}).

\bibitem{McDermott2009}
\bibinfo{author}{McDermott, R.}
\newblock \bibinfo{title}{Materials origins of decoherence in superconducting
  qubits}.
\newblock \emph{\bibinfo{journal}{IEEE Transactions on Applied
  Superconductivity}} \textbf{\bibinfo{volume}{19}}, \bibinfo{pages}{2--13}
  (\bibinfo{year}{2009}).

\bibitem{Oliver2013}
\bibinfo{author}{Oliver, W.~D.} \& \bibinfo{author}{Welander, P.~B.}
\newblock \bibinfo{title}{Materials in superconducting quantum bits}.
\newblock \emph{\bibinfo{journal}{MRS Bulletin}} \textbf{\bibinfo{volume}{38}},
  \bibinfo{pages}{816?825} (\bibinfo{year}{2013}).

\bibitem{Wenner2011}
\bibinfo{author}{Wenner, J.} \emph{et~al.}
\newblock \bibinfo{title}{Surface loss simulations of superconducting coplanar
  waveguide resonators}.
\newblock \emph{\bibinfo{journal}{Applied Physics Letters}}
  \textbf{\bibinfo{volume}{99}}, \bibinfo{pages}{113513}
  (\bibinfo{year}{2011}).

\bibitem{Wang2015}
\bibinfo{author}{Wang, C.} \emph{et~al.}
\newblock \bibinfo{title}{Surface participation and dielectric loss in
  superconducting qubits}.
\newblock \emph{\bibinfo{journal}{Applied Physics Letters}}
  \textbf{\bibinfo{volume}{107}}, \bibinfo{pages}{162601}
  (\bibinfo{year}{2015}).

\bibitem{Calusine2018}
\bibinfo{author}{Calusine, G.} \emph{et~al.}
\newblock \bibinfo{title}{Analysis and mitigation of interface losses in
  trenched superconducting coplanar waveguide resonators}.
\newblock \emph{\bibinfo{journal}{Applied Physics Letters}}
  \textbf{\bibinfo{volume}{112}}, \bibinfo{pages}{062601}
  (\bibinfo{year}{2018}).

\bibitem{Barends2010}
\bibinfo{author}{Barends, R.} \emph{et~al.}
\newblock \bibinfo{title}{Reduced frequency noise in superconducting
  resonators}.
\newblock \emph{\bibinfo{journal}{Applied Physics Letters}}
  \textbf{\bibinfo{volume}{97}}, \bibinfo{pages}{033507}
  (\bibinfo{year}{2010}).

\bibitem{Sage2011}
\bibinfo{author}{Sage, J.~M.}, \bibinfo{author}{Bolkhovsky, V.},
  \bibinfo{author}{Oliver, W.~D.}, \bibinfo{author}{Turek, B.} \&
  \bibinfo{author}{Welander, P.~B.}
\newblock \bibinfo{title}{Study of loss in superconducting coplanar waveguide
  resonators}.
\newblock \emph{\bibinfo{journal}{Journal of Applied Physics}}
  \textbf{\bibinfo{volume}{109}}, \bibinfo{pages}{063915}
  (\bibinfo{year}{2011}).

\bibitem{Megrant2012}
\bibinfo{author}{Megrant, A.} \emph{et~al.}
\newblock \bibinfo{title}{Planar superconducting resonators with internal
  quality factors above one million}.
\newblock \emph{\bibinfo{journal}{Applied Physics Letters}}
  \textbf{\bibinfo{volume}{100}}, \bibinfo{pages}{113510}
  (\bibinfo{year}{2012}).

\bibitem{Geerlings2012}
\bibinfo{author}{Geerlings, K.} \emph{et~al.}
\newblock \bibinfo{title}{Improving the quality factor of microwave compact
  resonators by optimizing their geometrical parameters}.
\newblock \emph{\bibinfo{journal}{Applied Physics Letters}}
  \textbf{\bibinfo{volume}{100}}, \bibinfo{pages}{192601}
  (\bibinfo{year}{2012}).

\bibitem{Sandberg2012}
\bibinfo{author}{Sandberg, M.} \emph{et~al.}
\newblock \bibinfo{title}{Etch induced microwave losses in titanium nitride
  superconducting resonators}.
\newblock \emph{\bibinfo{journal}{Applied Physics Letters}}
  \textbf{\bibinfo{volume}{100}}, \bibinfo{pages}{262605}
  (\bibinfo{year}{2012}).

\bibitem{Chang2013}
\bibinfo{author}{Chang, J.~B.} \emph{et~al.}
\newblock \bibinfo{title}{Improved superconducting qubit coherence using
  titanium nitride}.
\newblock \emph{\bibinfo{journal}{Applied Physics Letters}}
  \textbf{\bibinfo{volume}{103}}, \bibinfo{pages}{012602}
  (\bibinfo{year}{2013}).

\bibitem{Quintana2014}
\bibinfo{author}{Quintana, C.~M.} \emph{et~al.}
\newblock \bibinfo{title}{Characterization and reduction of
  microfabrication-induced decoherence in superconducting quantum circuits}.
\newblock \emph{\bibinfo{journal}{Applied Physics Letters}}
  \textbf{\bibinfo{volume}{105}}, \bibinfo{pages}{062601}
  (\bibinfo{year}{2014}).

\bibitem{Bruno2015}
\bibinfo{author}{Bruno, A.} \emph{et~al.}
\newblock \bibinfo{title}{Reducing intrinsic loss in superconducting resonators
  by surface treatment and deep etching of silicon substrates}.
\newblock \emph{\bibinfo{journal}{Applied Physics Letters}}
  \textbf{\bibinfo{volume}{106}}, \bibinfo{pages}{182601}
  (\bibinfo{year}{2015}).

\bibitem{Dunsworth2017}
\bibinfo{author}{Dunsworth, A.} \emph{et~al.}
\newblock \bibinfo{title}{Characterization and reduction of capacitive loss
  induced by sub-micron josephson junction fabrication in superconducting
  qubits}.
\newblock \emph{\bibinfo{journal}{Applied Physics Letters}}
  \textbf{\bibinfo{volume}{111}}, \bibinfo{pages}{022601}
  (\bibinfo{year}{2017}).

\bibitem{Leonard2018}
\bibinfo{author}{Leonard, E.} \emph{et~al.}
\newblock \bibinfo{title}{Digital coherent control of a superconducting qubit}.
\newblock \emph{\bibinfo{journal}{Phys. Rev. Applied}}
  \textbf{\bibinfo{volume}{11}}, \bibinfo{pages}{014009}
  (\bibinfo{year}{2019}).

\bibitem{OBrien2018}
\bibinfo{author}{O'Brien, K.} \emph{et~al.}
\newblock \bibinfo{title}{Scaling of variational quantum eigensolver
  performance in a superconducting quantum processor}.
\newblock \emph{\bibinfo{journal}{Bulletin of the American Physical Society}}
  \textbf{\bibinfo{volume}{63}} (\bibinfo{year}{2018}).

\bibitem{Koch2007}
\bibinfo{author}{Koch, J.} \emph{et~al.}
\newblock \bibinfo{title}{Charge-insensitive qubit design derived from the
  cooper pair box}.
\newblock \emph{\bibinfo{journal}{Phys. Rev. A}} \textbf{\bibinfo{volume}{76}},
  \bibinfo{pages}{042319} (\bibinfo{year}{2007}).

\bibitem{Gao2008}
\bibinfo{author}{Gao, J.} \emph{et~al.}
\newblock \bibinfo{title}{Experimental evidence for a surface distribution of
  two-level systems in superconducting lithographed microwave resonators}.
\newblock \emph{\bibinfo{journal}{Applied Physics Letters}}
  \textbf{\bibinfo{volume}{92}}, \bibinfo{pages}{152505}
  (\bibinfo{year}{2008}).

\bibitem{Kern1990}
\bibinfo{author}{Kern, W.}
\newblock \bibinfo{title}{The evolution of silicon wafer cleaning technology}.
\newblock \emph{\bibinfo{journal}{Journal of The Electrochemical Society}}
  \textbf{\bibinfo{volume}{137}}, \bibinfo{pages}{1887--1892}
  (\bibinfo{year}{1990}).

\bibitem{Vahidpour2017}
\bibinfo{author}{Vahidpour, M.} \emph{et~al.}
\newblock \bibinfo{title}{Superconducting through-silicon vias for quantum
  integrated circuits}.
\newblock \emph{\bibinfo{journal}{arXiv preprint arXiv:1708.02226}}
  (\bibinfo{year}{2017}).

\bibitem{Wisbey2010}
\bibinfo{author}{Wisbey, D.~S.} \emph{et~al.}
\newblock \bibinfo{title}{Effect of metal/substrate interfaces on
  radio-frequency loss in superconducting coplanar waveguides}.
\newblock \emph{\bibinfo{journal}{Journal of Applied Physics}}
  \textbf{\bibinfo{volume}{108}}, \bibinfo{pages}{093918}
  (\bibinfo{year}{2010}).

\bibitem{Fenner1989}
\bibinfo{author}{Fenner, D.~B.}, \bibinfo{author}{Biegelsen, D.~K.} \&
  \bibinfo{author}{Bringans, R.~D.}
\newblock \bibinfo{title}{Silicon surface passivation by hydrogen termination:
  A comparative study of preparation methods}.
\newblock \emph{\bibinfo{journal}{Journal of Applied Physics}}
  \textbf{\bibinfo{volume}{66}}, \bibinfo{pages}{419--424}
  (\bibinfo{year}{1989}).

\bibitem{morita1990}
\bibinfo{author}{Morita, M.}, \bibinfo{author}{Ohmi, T.},
  \bibinfo{author}{Hasegawa, E.}, \bibinfo{author}{Kawakami, M.} \&
  \bibinfo{author}{Ohwada, M.}
\newblock \bibinfo{title}{Growth of native oxide on a silicon surface}.
\newblock \emph{\bibinfo{journal}{Journal of Applied Physics}}
  \textbf{\bibinfo{volume}{68}}, \bibinfo{pages}{1272--1281}
  (\bibinfo{year}{1990}).

\bibitem{Woods2018}
\bibinfo{author}{Woods, W.} \emph{et~al.}
\newblock \bibinfo{title}{Determining interface dielectric losses in
  superconducting coplanar waveguide resonators}.
\newblock \emph{\bibinfo{journal}{arXiv preprint arXiv:1808.10347}}
  (\bibinfo{year}{2018}).

\bibitem{Gambetta2017}
\bibinfo{author}{Gambetta, J.~M.}, \bibinfo{author}{Chow, J.~M.} \&
  \bibinfo{author}{Steffen, M.}
\newblock \bibinfo{title}{Building logical qubits in a superconducting quantum
  computing system}.
\newblock \emph{\bibinfo{journal}{npj Quantum Information}}
  \textbf{\bibinfo{volume}{3}}, \bibinfo{pages}{2} (\bibinfo{year}{2017}).

\bibitem{Vissers2012}
\bibinfo{author}{Vissers, M.~R.}, \bibinfo{author}{Kline, J.~S.},
  \bibinfo{author}{Gao, J.}, \bibinfo{author}{Wisbey, D.~S.} \&
  \bibinfo{author}{Pappas, D.~P.}
\newblock \bibinfo{title}{Reduced microwave loss in trenched superconducting
  coplanar waveguides}.
\newblock \emph{\bibinfo{journal}{Applied Physics Letters}}
  \textbf{\bibinfo{volume}{100}}, \bibinfo{pages}{082602}
  (\bibinfo{year}{2012}).

\bibitem{Martinis2005}
\bibinfo{author}{Martinis, J.~M.} \emph{et~al.}
\newblock \bibinfo{title}{Decoherence in josephson qubits from dielectric
  loss}.
\newblock \emph{\bibinfo{journal}{Phys. Rev. Lett.}}
  \textbf{\bibinfo{volume}{95}}, \bibinfo{pages}{210503}
  (\bibinfo{year}{2005}).

\bibitem{Barends2011}
\bibinfo{author}{Barends, R.} \emph{et~al.}
\newblock \bibinfo{title}{Minimizing quasiparticle generation from stray
  infrared light in superconducting quantum circuits}.
\newblock \emph{\bibinfo{journal}{Applied Physics Letters}}
  \textbf{\bibinfo{volume}{99}}, \bibinfo{pages}{113507}
  (\bibinfo{year}{2011}).

\end{thebibliography}

%%%%%%%%%% Merge with supplemental materials %%%%%%%%%%
\pagebreak
\widetext
% \begin{center}
% \textbf{\large Supplementary Materials}
% \end{center}

\begin{titlepage}
	\centering
	{\large \textbf{Manufacturing low dissipation superconducting quantum processors} \\ \normalfont Supplementary Materials \par}

\end{titlepage}
%%%%%%%%%% Merge with supplemental materials %%%%%%%%%%
%%%%%%%%%% Prefix a "S" to all equations, figures, tables and reset the counter %%%%%%%%%%
\setcounter{equation}{0}
\setcounter{figure}{0}
\setcounter{table}{0}
\setcounter{section}{0}
\setcounter{page}{1}
\makeatletter
\renewcommand{\theequation}{S\arabic{equation}}
\renewcommand{\thefigure}{S\arabic{figure}}
\renewcommand{\thetable}{S\Roman{table}}
\renewcommand{\thetable}{S\Roman{section}}
\renewcommand{\bibnumfmt}[1]{[S#1]}
\renewcommand{\citenumfont}[1]{S#1}

\onecolumngrid

\section{Device design and measurement setup}
\label{supplement}

\begin{figure}[ht]
\centering
\includegraphics[width=0.375\columnwidth]{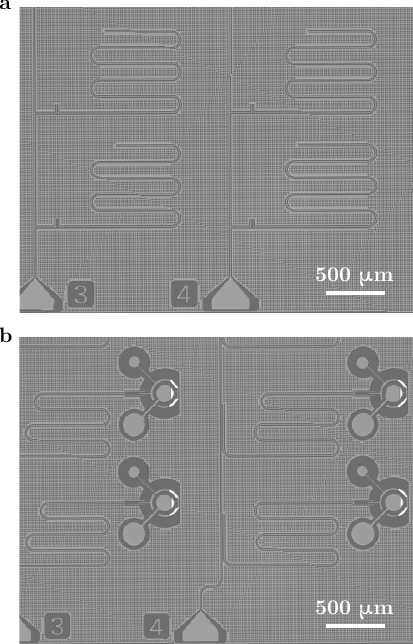}
\caption{\textbf{Device design.} \textbf{a,} Optical image of a resonator device. \textbf{b,} Optical image of a qubit device.} 
\label{fig:design}
\end{figure}

To enable fast iteration of characterizing each fabrication technique as well as the reproducibility of their effects, we test two separate device designs. The first design is used to measure internal quality factors of resonators (Fig.~\ref{fig:design}a). It consists of two parallel coplanar transmission lines each capacitively coupled to a set of four CPWRs. The resonator frequency band is between $5.2\,$GHz and $5.6\,$GHz with $50$ - $100\,$MHz detuning between each pair. An optical image of the lower half of a device with this design is shown in Fig.~~\ref{fig:design}a. 

The second design is used to characterize relaxation times of the superconducting qubits capacitively coupled to CPWRs (Fig.~\ref{fig:design}b). Similar to the first design, two sets of four CPWRs are coupled to two parallel transmission lines. However, the coupling is inductive in this case. Qubit frequencies vary between $3.8\,$GHz and $4.2\,$GHz with a neighbouring qubit nominal detuning of $50\,$MHz. A micrograph of the lower half of a device with this second design is shown in Fig.~\ref{fig:design}b.

To test the performance of these devices, we mount three $5.3\times5.3\,$mm$^2$ samples into a holder thermally anchored to the bottom plate of a dilution refrigerator with base temperature of $10\,$mK. The frequency response of the CPWRs is acquired with a vector network analyzer (VNA).
The full attenuation of the input line, from the VNA to the sample holder, is characterized beforehand at base temperature. The measured transmission coefficient is fitted with a theoretical single-pole model in the complex plane and the input driving power is converted to an average photon number~\cite{Bruno2015}. The driving power is then adjusted to reach single photon driving. We use Ettus Research USRP X300 to measure the relaxation times of the eight qubits on each device typically $10-15$ continuous hours.

The sample holder is shielded from external magnetic fields by one superconducting and two cryoperm cans. An extra internal copper can is coated with a blackbody absorber to suppress stray infrared radiations~\cite{Barends2011}. The input microwave signal is attenuated by a series of bulk SMA attenuators anchored at different temperature stages and further filtered by a $7.65\,$GHz low-pass filter. The input coaxial line has a total attenuation of $76\,$dB at DC and room temperature. The output signal from the sample is filtered by two isolators before being amplified by a high electron mobility transistor at $4\,$K and a series of room temperature amplifiers.

\section{Details of fabrication procedures}

All devices are fabricated on high-resistivity Si wafers ($\rho \geq 10$~k$\Omega$ cm). Prior to metallization, Si wafers are cleaned following Ref.~\cite{Kern1990} and subsequently immersed in BOE 5:1 solution for two minutes to remove the native silicon-oxide. Ar$^{+}$ ion milling is conducted for two minutes at 600~V and 118~mA, and HMDS passivation for ten minutes at 120$^\circ$C hot plate. After substrate treatment, wafers are coated with $200\,$nm of Nb ($99.999\,$\% purity) using PVD at $350\,$W power.

Large device features are defined with optical lithography using either $1.5\,\mu$m of positive AZ3318D resist (for anisotropic etch) or $7.3\,\mu$m of positive AZ9260 resist (for isotropic and tapered etch). RIE is performed with SF$_6$ for 3-5.5 minutes at 10-200~mTorr and 100-150~W. Oxygen plasma ashing is conducted for two minutes at 150~W at a pressure of 12~mTorr.

The JJ definition process consists of patterning a bilayer of MMA and PMMA resists with electron beam lithography, followed by $\pm$40$^\circ$ double-angle evaporation of $30\,$nm and $50\,$nm of first and second Al layers. For the bandage layer, an additional electron beam lithography step is used to lift off $250\,$nm of Al following four minutes of ion milling at $200\,$V or $400\,$V and $30\,$mA.

\section{Additional data}
\begin{figure*}[ht]
\centering
\includegraphics[width=0.875\columnwidth]{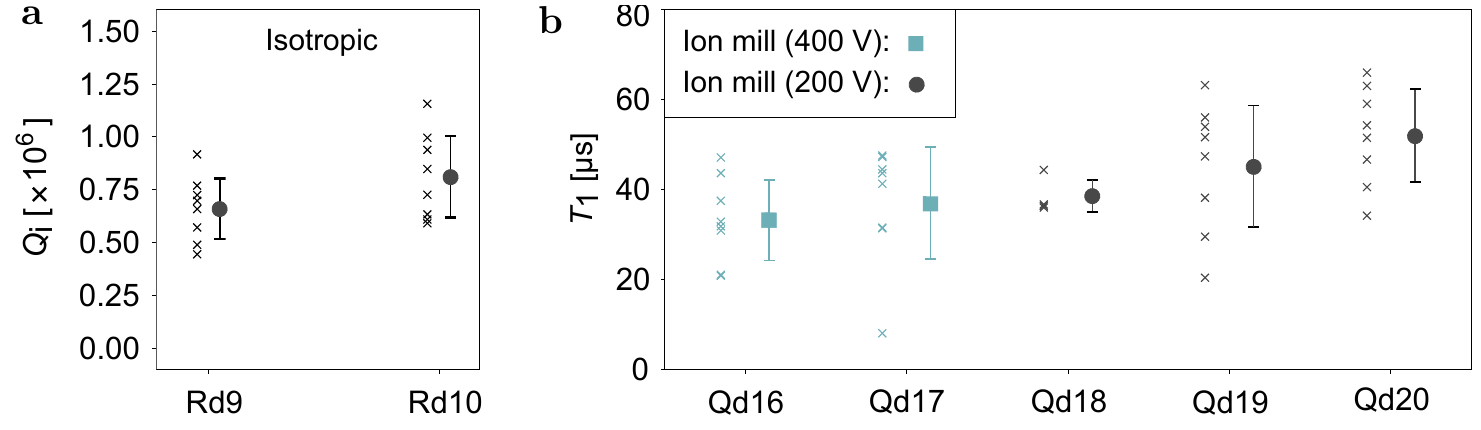}
\caption{\textbf{Quality factors and relaxation times of additional devices with control MS interface treatment.} \textbf{a,} $Q_\textrm{i}$'s at single photon powers of resonator devices with isotropic etching. \textbf{b,}  $T_1$'s of qubit devices with tapered etch profile and an additional bandage layer, grouped by the ion milling voltage used before the application of the bandage layer: $400\,$V and $200\,$V.}
\label{fig:ba_no_hmds}
\end{figure*}

In Fig.~\ref{fig:ba_no_hmds}a we show some complimentary CPWR quality factors from devices that received control substrate treatment and had isotropic etch profile, and in Fig.~\ref{fig:ba_no_hmds}b some qubit relaxation times from additional devices with the same substrate treatment having tapered etch profile and bandage layer to connect the JJs.

\end{document}